\documentclass[a4paper]{jpconf}
\usepackage{graphicx}
\usepackage{amsmath}
\usepackage[amssymb, mediumspace]{SIunits}
\usepackage{multicol}
\usepackage[
  font={small,sl},
  labelfont={sc,bf},
  labelsep={space},
  format={plain},
  justification={justified},
  singlelinecheck={yes},
  figureposition={bottom},
  tableposition={top},
  margin={4em}
  ]{caption}[2008/04/01]
\usepackage{color}
  \definecolor{my_color_for_extrainfo}{rgb}{0.4,0.2,0.2}
  \definecolor{myred}{rgb}{0.9,0.1,0.1}
  \definecolor{myDarkRed}{rgb}{0.5,0,0}       
  \definecolor{myDarkBlue}{rgb}{0,0,0.6}      
  \definecolor{myDarkGreen}{rgb}{0,0.4,0}     
  \definecolor{myDarkGray}{rgb}{0.4,0.4,0.4}  
  \definecolor{black}{rgb}{0,0,0}
  \definecolor{white}{rgb}{1,1,1}
\usepackage{ifthen}
  \newcommand{\showextrainfo}{1}
  \newcommand{\extrainfo}[1]{%
    {\color{my_color_for_extrainfo}%
      \ifthenelse{\isundefined{\showextrainfo}}{}{#1}%
    }%
  }

  \newcommand{\inred}[1]{{#1}}

  \newcommand{\figurefolder}[0]{}
  
\usepackage[
  a4paper={true},
  colorlinks={true},
  linkcolor={myDarkBlue},
  citecolor={myDarkGreen},
  urlcolor={myDarkBlue},
  hyperfootnotes={false}
  ]{hyperref}

\newcommand{\pderiv}[2]{\frac{\partial #1}{\partial #2}} 

\begin{document}

\title{Coupling JOREK and STARWALL Codes for Non-linear Resistive-wall Simulations}

\author{M.~H\"olzl$^1$, P.~Merkel$^1$, G.T.A.~Huysmans$^2$, E.~Nardon$^3$, E.~Strumberger$^1$, R.~McAdams$^{4,5}$, I.~Chapman$^5$, S.~G\"unter$^1$, K.~Lackner$^1$}

\address{
$^1$Max-Planck-Institute for Plasmaphysics, EURATOM Association,
  Boltzmannstr. 2, 85748 Garching, Germany \\
$^2$ITER Organisation, Route de Vinon sur Verdon,
  St-Paul-lez-Durance, France \\
$^3$CEA, IRFM, CEA Cadarache, F-13108 St Paul-lez-Durance, France \\
$^4$York Plasma Institute, University of York, York, YO10 5DD, UK \\
$^5$Euratom/CCFE Fusion Association, Culham Science Centre, Abingdon, OX14 3DB, UK
}

\ead{mhoelzl@ipp.mpg.de}

\inred{
\begin{abstract}
The implementation of a resistive-wall extension to the non-linear MHD-code
JOREK via a coupling to the vacuum-field code STARWALL is presented along with
first applications and benchmark results. Also, non-linear saturation in the
presence of a resistive wall is demonstrated.
After completion of the ongoing verification process, this code extension
will allow to perform non-linear simulations of MHD instabilities in the
presence of three-dimensional resistive walls with holes for limited and X-point
plasmas.
\end{abstract}
}

\section{Introduction}

Plasma instabilities growing on fast time-scales such as vertical displacement events,
disruptions, external kink modes, or edge localized modes are connected with
a time-dependent magnetic perturbation outside the plasma. The mirror currents
induced in conducting
structures around the plasma\footnote{For simplicity, all conducting structures are denoted wall
in the following.} by the varying field perturbations act
back onto the instabilities. Thus, linear and
non-linear dynamics of the plasma can be strongly influenced by the wall. An
external kink mode \inred{driven by strong edge current
densities}, for instance, is converted
into a resistive wall mode~(RWM) in the presence of a conducting wall close to the
plasma edge and grows on the timescale of resistive wall-current decay as described
in great detail in Refs.~\cite{Chu2010,Igochine2012}.

We describe the status of a resistive wall extension for
the non-linear MHD code JOREK~\cite{Czarny2008,Huysmans2010} as well as first
benchmark results. The implementation is done by coupling JOREK
to a modified version of the STARWALL code~\cite{Merkel2006} which
determines the magnetic field in the vacuum region in the presence of
three-dimensional walls with holes. The ``response'' of conducting structures
to magnetic field perturbations is represented by response-matrices
computed by STARWALL. These matrices are used in JOREK
to keep track of wall-currents and calculate the tangential magnetic field required
for the modified boundary conditions as discussed in Section~\ref{:impl:boundint}.

The article provides some information about the
STARWALL code in Section~\ref{:starwall}. The coupling equations
are derived in Section~\ref{:impl}.
First benchmarks of the resistive wall extension are
presented in Section~\ref{:res}. Conclusions and a brief outlook are given in
Section~\ref{:concl}.

\section{The STARWALL Code}\label{:starwall}

STARWALL solves the vacuum magnetic field equation outside the JOREK
computational domain in presence of a three-dimensional conducting wall with
holes as a Neumann-like problem. The continuity of the magnetic field component
\inred{normal} to the boundary of the JOREK computational domain (called interface)
is used as a boundary condition. The wall is represented by infinitely thin
triangles which allows to approximate realistic wall structures very well using
an effective resistance $\eta_\text{w}/d_\text{w}$ where $\eta_\text{w}$
and $d_\text{w}$ denote the resistivity of the wall material\footnote{Wall
resistivity is normalized the same way as plasma resistivity
(see Ref.~\cite{Hoelzl2012A} for JOREK normalizations).} and its thickness,
respectively. Wall currents are assumed constant within each triangle
such that they can be described by \inred{current potentials $Y_k$ at the
triangle nodes}~\cite{Merkel2006}\footnote{For simplicity, we will (inaccurately) speak about
wall currents when wall current potentials are referred to.}.

In case of an ideally conducting wall, solving the vacuum field equation
results in an algebraic expression for the magnetic field component
tangential to the interface\footnote{$B_{\texttt{tan}}$ is the component
of $(\mathbf{B}-\mathbf{B}_\phi)$ tangential to the interface, where $\mathbf{B}_\phi$
denotes the toroidal field component.}
in terms of the \inred{normal} component. Wall currents
don't need to be considered explicitly as they are instantaneously given by the
\inred{normal} field at the interface. Using the
ideal wall response-matrix $\hat{M}^\text{id}$
calculated by STARWALL, this can be written
in the following way\footnote{For practical
reasons, the algebraic expression is written in terms of the poloidal flux $\Psi$
instead of the \inred{normal} field component (which can be calculated
from $\Psi$, of course).}:
\begin{equation}\label{eq:freebound80}
  B_{\texttt{tan}}
  = \sum_i b_i\;B_{\texttt{tan},i}
  = \sum_i b_i
      \sum_j\hat{M}^\text{id}_{i,j}\;\Psi_j.
\end{equation}
Here, $b_i$ denotes a JOREK basis function consisting of a \inred{1D Bezier basis
function along the poloidal direction on the interface} and Fourier basis
function in toroidal
direction. $\Psi_j$ denotes poloidal flux coefficients.
\inred{%
The poloidal flux at the interface is given by $\Psi=\sum_j b_j\;\Psi_j.$
}

When a resistive wall is considered, wall currents cannot be
eliminated anymore. In this case, the tangential magnetic
field is given by
\begin{equation}\label{eq:freebound100}
  B_{\texttt{tan}}
  = \sum_i b_i\left(
      \sum_j\hat{M}^\text{ee}_{i,j}\;\Psi_j
      + \sum_k\hat{M}^\text{ey}_{i,k}\;Y_k
    \right),
\end{equation}
where wall currents evolve in time according to
\begin{equation}\label{eq:freebound200}
  \dot{Y}_k =
    - \frac{\eta_\text{w}}{d_\text{w}}\;\hat{M}^\text{yy}_{k,k}\;Y_k
    - \sum_j\hat{M}^\text{ye}_{k,j}\;\dot{\Psi}_j.
\end{equation}
Here, $\hat{M^\text{ee}}$, $\hat{M^\text{ey}}$, $\hat{M^\text{ye}}$, and $\hat{M^\text{yy}}$
denote resistive response matrices determined by
STARWALL. Indices $i$ and $j$ run over all boundary degrees of freedom of the respective
variable and $k$ over all wall current potentials. For cross-checking, a relation
between ideal and resistive response matrices can be derived by letting
$\eta_\text{w}\rightarrow\infty$ (\ref{::app1}).

Information about the STARWALL code coupled with CASTOR (sometimes called
STARWALL\_C) can be found in Refs.~\cite{Merkel2006,Strumberger2011A}. It is similar to the
STARWALL code coupled with JOREK (sometimes called STARWALL\_J).
An article describing more details is in preparation by the
STARWALL author Peter Merkel.

\section{Implementation in JOREK}\label{:impl}

For the coupling of JOREK and STARWALL, Eqs.~\eqref{eq:freebound100}
and~\eqref{eq:freebound200} are discretized in time as described in
Section~\ref{:impl:discret}. The vacuum response then enters into a
natural boundary condition in JOREK as discussed in Section~\ref{:impl:boundint}.

\subsection{Time-Discretization}\label{:impl:discret}

Equation~\eqref{eq:freebound100} is
evaluated at the new timestep $n+1$ and discretizations
$\Psi^{n+1}   = \Psi^{n}   + \delta \Psi^{n}$ and
$Y^{n+1}      = Y^{n}      + \delta Y^{n}$
are used, where superscripts indicate evaluation at the given timestep.
The tangential magnetic field is thus given by
\begin{equation} \label{eq:freebound400}
  B_{\texttt{tan}}^{n+1}
  = \sum_i b_i\left[
      \sum_j\hat{M}^\text{ee}_{i,j}\;\left(\Psi_j^n+\delta\Psi_j^n\right)
      + \sum_k\hat{M}^\text{ey}_{i,k}\; \left(Y_k^n + \delta Y_k^n\right)
    \right].
\end{equation}

\inred{
The general time-evolution scheme described in Ref.~\cite{HIRSCH1989},
\begin{equation}\label{eq:timestep2}
    \left[(1+\xi)\left(\pderiv{\mathbf{X}}{\mathbf{u}}\right)^n-\Delta t\theta\left(\pderiv{\mathbf{Z}}{\mathbf{u}}\right)^n\right]\delta\mathbf{u}^n
      =\Delta t\;\mathbf{Z}^n+\xi\left(\pderiv{\mathbf{X}}{\mathbf{u}}\right)^{n-1}\delta\mathbf{u}^{n-1},
\end{equation}
is used for the JOREK equations written in the form
$\partial\mathbf{X(\mathbf{u})}/\partial t=\mathbf{Z(\mathbf{u})}$, where
$\mathbf{u}$ denotes the vector of JOREK physical variables like temperature
or density, $\mathbf{X}$ and $\mathbf{Z}$ the left- respectively right-hand
side terms of the time-evolution equations, and $\theta$ and $\xi$ are numerical
parameters\footnote{Parameter values satisfying $\theta-\xi=0.5$ are required to guarantee
second-order accuracy of the time-evolution scheme. For instance, $(\theta=0.5, \xi=0)$
corresponds to a linearized Crank-Nicholson scheme and $(\theta=1, \xi=0.5)$ to the linearized
Gears scheme.}. The
same general time-evolution method needs to be
applied to the wall-current evolution (Eq.~\eqref{eq:freebound200}) giving
}
\begin{equation}\begin{split}\label{eq:freebound500}
  &(1+\xi)\left[\delta Y^n_k+\sum_j\hat{M}^\text{ye}_{k,j}\;\delta\Psi^n_j\right]
  + \Delta t\;\theta\;\frac{\eta_\text{w}}{d_\text{w}}\;\hat{M}^\text{yy}_{k,k}\;\delta Y^n_k \\
  = &-\Delta t\;\frac{\eta_\text{w}}{d_\text{w}}\;\hat{M}^\text{yy}_{k,k}\;Y^n_k
  +\xi\left[\delta Y^{n-1}_k+\sum_j\hat{M}^\text{ye}_{k,j}\;\delta\Psi^{n-1}_j\right].
\end{split}\end{equation}
Terms with $\delta Y_k^n$ are brought to the left hand side.
\inred{%
The equation then reads
\begin{equation}\begin{split}
&\left(1+\xi+\Delta t\theta\frac{\eta_\text{w}}{d_\text{w}}\hat{M}^\text{yy}_{k,k}\right)\delta Y^n_k \\
  = &-(1+\xi)\sum_j\hat{M}^\text{ye}_{k,j}\;\delta\Psi^n_j
    -\Delta t\frac{\eta_\text{w}}{d_\text{w}}\hat{M}^\text{yy}_{k,k}Y^n_k
    +\xi\delta Y^{n-1}_k
    +\xi\sum_j\hat{M}^\text{ye}_{k,j}\;\delta\Psi^{n-1}_j.
\end{split}\end{equation}
}%
After solving for $\delta Y_k^n$, one gets
\begin{equation}\label{eq:freebound600}
  \delta Y_k^n = \sum_j \hat{A}_{k,j}\;\delta\Psi^n_j
    + \hat{B}_{k,k}\;Y^n_k
    + \hat{C}_{k,k}\;\delta Y^{n-1}_k
    + \sum_j\hat{D}_{k,j}\;\delta\Psi^{n-1}_j,
\end{equation}
where some of the ``derived response matrices''
\begin{equation}
\begin{array}{lll}
  \hat{S}_{k,k} = 1+\xi+\Delta t\theta\frac{\eta_\text{w}}{d_\text{w}}\hat{M}^\text{yy}_{k,k}
  ~~~~~& \hat{D}_{k,j} = \xi\hat{M}^\text{ye}_{k,j} / \hat{S}_{k,k}
  ~~~~~& \hat{H}_{i,j} = \hat{M}^\text{ee}_{i,j} \label{eq:freebound630}
  \\
  \hat{A}_{k,j} = -(1+\xi)\;\hat{M}^\text{ye}_{k,j} / \hat{S}_{k,k}
  ~~~~~& \hat{E}_{i,j} = \hat{M}^\text{ee}_{i,j}+\sum_k\hat{M}^\text{ey}_{i,k}\;\hat{A}_{k,j}
  ~~~~~& \hat{J}_{i,j} = \sum_k\hat{M}^\text{ey}_{i,k}\;\hat{D}_{k,j}
  \\
  \hat{B}_{k,k} = -\Delta t\frac{\eta_\text{w}}{d_\text{w}}\hat{M}^\text{yy}_{k,k} / \hat{S}_{k,k}
  ~~~~~& \hat{F}_{i,k} = \hat{M}^\text{ey}_{i,k}(1+\hat{B}_{k,k})
  ~~~~~& \hat{K}_{k,l} = -\Delta t\hat{M}^\text{yy}_{k,k}\hat{S}_{l,k}
  \\
  \hat{C}_{k,k} = \xi / \hat{S}_{k,k}
  ~~~~~& \hat{G}_{i,k} = \hat{M}^\text{ey}_{i,k}\;\hat{C}_{k,k}
  ~~~~~& \hat{L}_{i,l} = \sum_k\hat{M}^\text{ey}_{i,k}\;\hat{K}_{k,l}
\end{array}
\end{equation}
have been used.
\inred{%
These matrices are computed at the beginning of a JOREK simulation from
the STARWALL response matrices. They need to be updated if one or more parameters
entering the definitions have changed, e.g., $\Delta t$, $\xi$, or $\eta_\text{w}$.
}%
Plugging Eq.~\eqref{eq:freebound600} into Eq.~\eqref{eq:freebound400}
\inred{%
gives
\begin{equation}\begin{split}
  B^{n+1}_\text{tan}=\sum_i b_i\Biggl[
    &\sum_j\left(\hat{M}^\text{ee}_{i,j}+\sum_k\hat{M}^\text{ey}_{i,k}\;\hat{A}_{k,j}\right)\delta\Psi^n_j
    +\sum_k\hat{M}^\text{ey}_{i,k}\left(1+\hat{B}_{k,k}\right)Y^n_k \\
    &+\sum_k\hat{M}^\text{ey}_{i,k}\hat{C}_{k,k}\;\delta Y^{n-1}_k
    +\sum_j\hat{M}^\text{ee}_{i,j}\Psi^n_j
    +\sum_j\sum_k\hat{M}^\text{ey}_{i,k}\hat{D}_{k,j}\delta\Psi^{n-1}_j
  \Biggr]
\end{split}\end{equation}
}%
and, making use of Eq.~\eqref{eq:freebound630}, we get
\begin{equation}\label{eq:freebound700}
  B_{\texttt{tan}}^{n+1} = \sum_i b_i \left[
    \sum_j\hat{E}_{i,j}\;\delta\Psi^n_j
    +\sum_k\hat{F}_{i,k}\;Y^n_k
    +\sum_k\hat{G}_{i,k}\;\delta Y^{n-1}_k
    +\sum_j\hat{H}_{i,j}\;\Psi^n_j
    +\sum_j\hat{J}_{i,j}\;\delta\Psi^{n-1}_j
  \right].
\end{equation}

\subsection{Boundary Integral}\label{:impl:boundint}

The fixed boundary conditions for poloidal flux and plasma current corresponding
to an ideally conducting wall are removed from JOREK. As a consequence, a boundary integral in the
current definition equation resulting from partial integration (which vanishes in
fixed-boundary JOREK) needs to be considered now. It
can be written in terms of the tangential magnetic field such that we can insert
Eq.~\eqref{eq:freebound700} as a
natural boundary condition. Details are given in the following.

The current definition equation, $j=\Delta^*\Psi$, is written in weak form as
\begin{equation}\label{eq:freebound900}
  \int dV \frac{j^*_l}{R^2}\;\left(j - \Delta^*\Psi\right) = 0.
\end{equation}
\inred{where $j^*_l$ denotes the test-functions taken to be identical with the basis-functions.}
Using $\Delta^*\Psi\equiv R^2\;\nabla\cdot(R^{-2}\;\nabla\Psi)$, we get
\begin{equation}\label{eq:freebound1000}
  \int dV\;\frac{j^*_l}{R^2}\;j - \int dV\;j^*_l\;\nabla\cdot\left(\frac{1}{R^2}\nabla\Psi\right) = 0,
\end{equation}
where the second term of the resulting expression can be integrated by parts
\inred{%
($\int dV\;a\;\nabla\cdot\mathbf{b} =
     - \int dV\;\nabla a\cdot\mathbf{b} 
     + \oint dA\;a\;\mathbf{b}\cdot\hat{\mathbf{n}}$)
}
yielding
\begin{equation}\label{eq:freebound1100}
  \int dV\;\frac{j^*_l}{R^2}\;j
  + \int dV\;\frac{1}{R^2}\;\nabla j^*_l\cdot\nabla\Psi
  - \oint dA\;\frac{j^*_l}{R}\;\underbrace{\left(\nabla\Psi\cdot \hat{\mathbf{n}}/R\right)}_{\equiv B_{\texttt{tan}}} = 0.
\end{equation}
Here, $\hat{n}$ denotes the unit vector normal to the interface and the tangential
field is identified in the boundary integral \inred{(refer to~\ref{::app2} for details)}.
Eq.~\eqref{eq:freebound700} is inserted into Eq.~\eqref{eq:freebound1100} and after
separating implicit and explicit terms, we have derived the form of the current
equation implemented in JOREK,
\begin{equation}\begin{split}\label{eq:freebound1600}
  &\sum_{i_\text{elem}}
  \int\frac{dV}{R^2}\left(j^*_l\;\delta j^n+\nabla j^*_l\cdot\nabla\delta\Psi^n\right)
  - \sum_{i_\text{bnd}}\oint\;dA\frac{j^*_l}{R}\sum_i b_i\sum_j \hat{E}_{i,j}\;\delta\Psi_j^n \\
  = &- \sum_{i_\text{elem}}\int\frac{dV}{R^2}\left(j^*_l\;j^n+\nabla j^*_l\cdot\nabla\Psi^n\right) \\
  &+ \sum_{i_\text{bnd}}\oint\;dA\frac{j^*_l}{R}
      \sum_i b_i\left[
        \sum_k\left(\hat{F}_{i,k}\;Y^n_k+\hat{G}_{i,k}\;\delta Y^{n-1}_k\right)
        +\sum_j\left(\hat{H}_{i,j}\;\Psi^n_j+\hat{J}_{i,j}\;\delta\Psi^{n-1}_j\right)\right],
\end{split}\end{equation}
with
\begin{equation}
  \int  dV=\sum_{i_\text{elem}}\int du\;d\inred{v}\;d\phi\;J_2\;R,~~~\text{and}~~~
  \oint dA=\sum_{i_\text{bnd}}\int du\;d\phi\;R\sqrt{(R_{,u})^2 + (Z_{,u})^2},
\end{equation}
\inred{where the integral over $u$ and $v$ is carried out by Gauss-quadrature:
$\int du\;X(u) = \sum_{m_G} w_{m_G}\;X(u_{m_G})$ with weights $w_{m_G}$ and positions $u_{m_G}$.
Coordinate $u$ corresponds to the element-local coordinate of the 2D Bezier elements oriented
poloidally along the interface. The other element-local coordinate is denoted $v$.}
$J_2$ denotes the 2D-Jacobian in the poloidal plane.
Wall-currents are updated after each time-step according to
Eq.~\eqref{eq:freebound600} to guarantee consistency with
the implicit time-stepping of JOREK.

\begin{figure}
\centering
  \includegraphics[width=0.48\textwidth]{\figurefolder 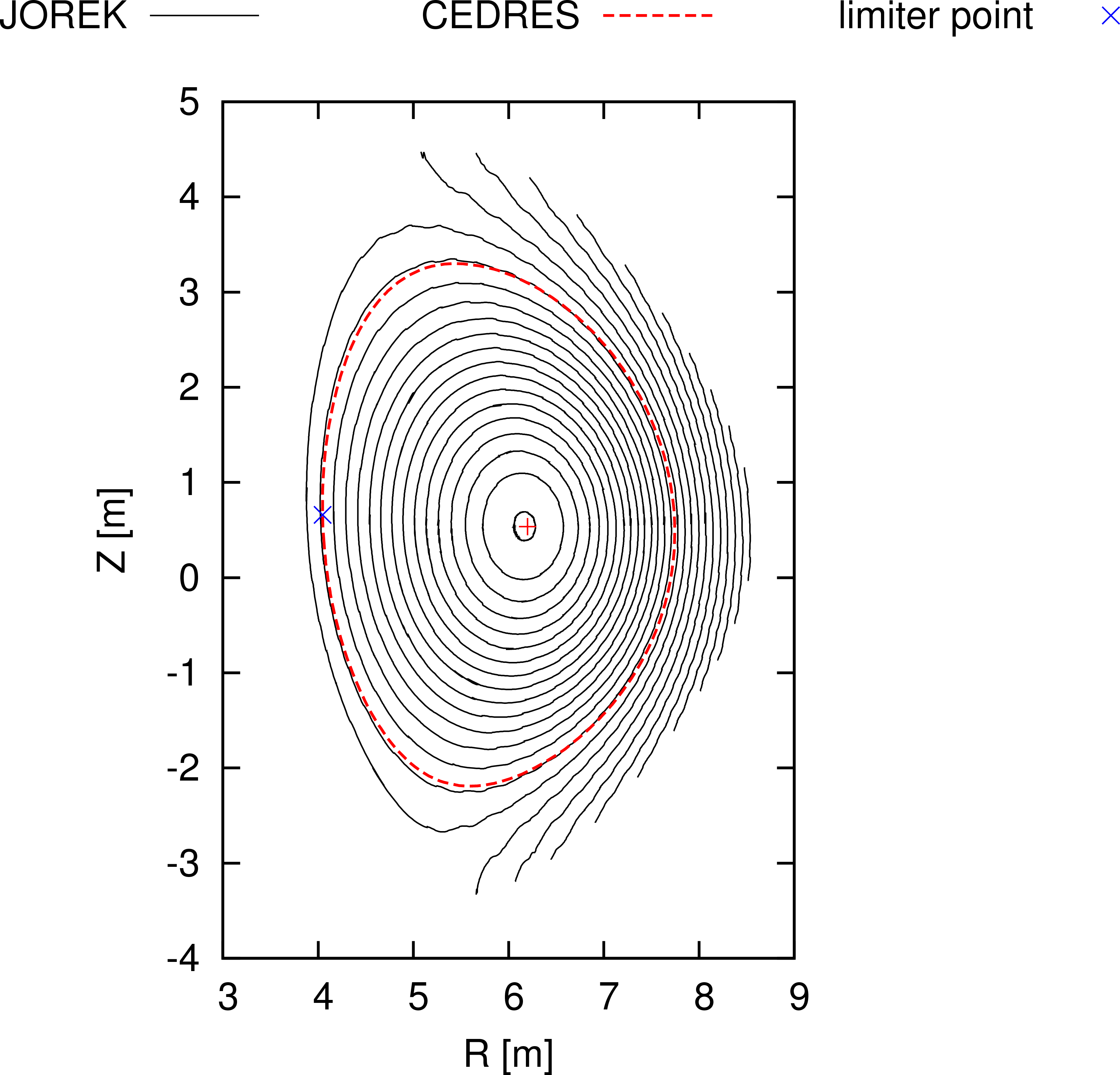}
  \includegraphics[width=0.48\textwidth]{\figurefolder 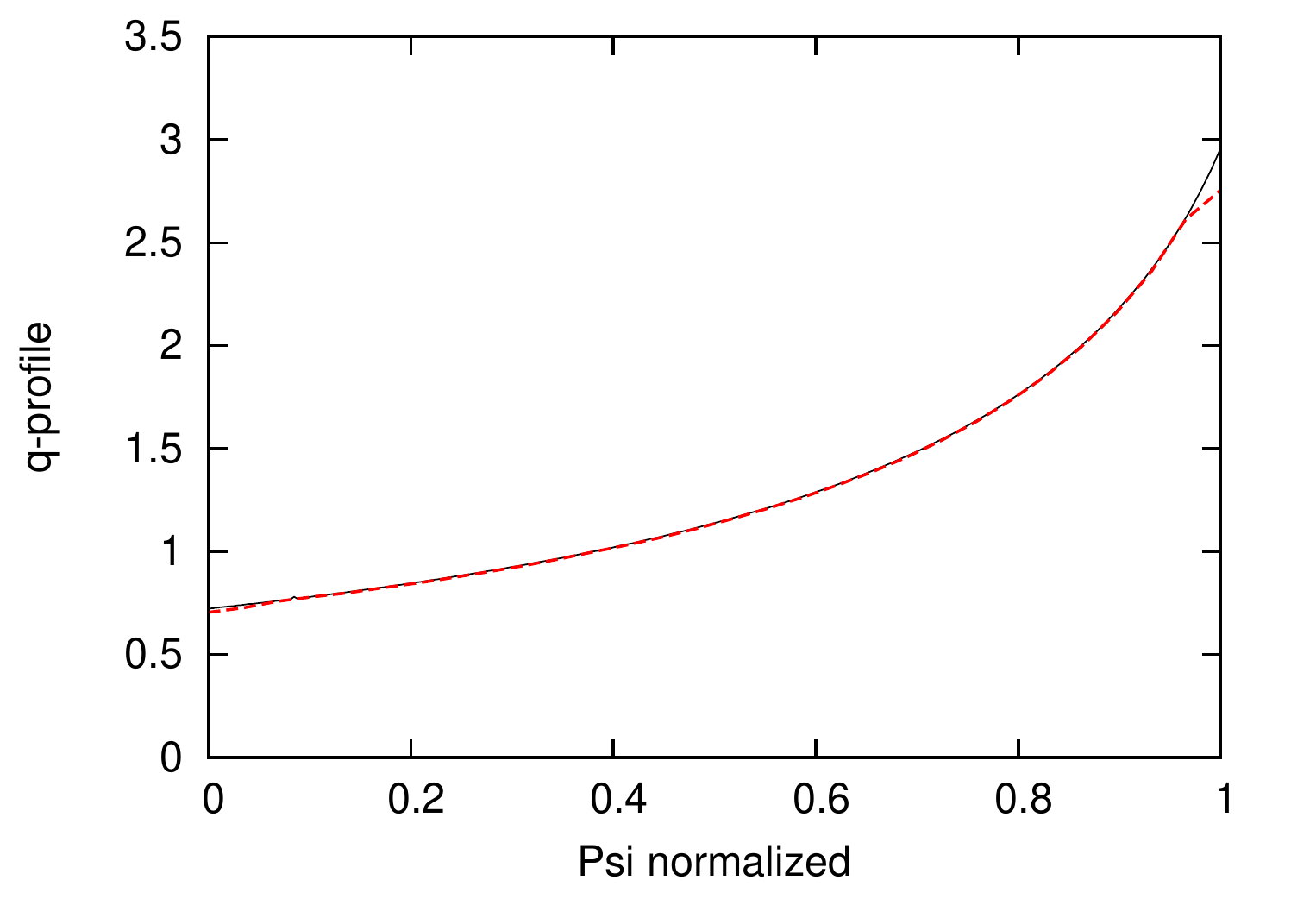}
\caption{
  \newline \textbf{Left:} JOREK flux-surfaces of a limiter equilibrium (limiter point
  shown as blue cross) are compared to CEDRES++ last closed flux-surface and magnetic axis.
  \newline \textbf{Right:} Comparison of the respective q-profiles.}
\label{fig:equil-compare}
\end{figure}

For the $n=0$ component, poloidal field coils need to be taken into account
properly. This is done by replacing the expression for the tangential magnetic field,
Eq.~\eqref{eq:freebound100}, by
\begin{equation}\label{eq:freebound2000}
  B_{\texttt{tan}}
  = \sum_i b_i\left(
      \sum_j\hat{M}^\text{ee}_{i,j}\;(\Psi_j-\Psi^\text{coil}_j)
      + \sum_k\hat{M}^\text{ey}_{i,k}\; Y_k
      + B^\text{coil}_{\texttt{tan},i}
    \right).
\end{equation}
Here, $\Psi^\text{coil}_j$ and $B^\text{coil}_{\texttt{tan},i}$ denote time-independent
coil contributions to poloidal flux and tangential field at the interface,
respectively. This leads to two additional terms in Eq.~\eqref{eq:freebound1600}.
A similar boundary integral also occurs in the Grad-Shafranov equation giving the
plasma equilibrium. Here, only the $n=0$ component needs to be considered
(axisymmetric equilibria in JOREK) and no time-discretization is required.

\section{First Benchmarks and Results}\label{:res}

In this Section, we present first results of the ongoing benchmark effort for code
validation.

\subsection{Free-boundary equilibrium}

Computing a free-boundary equilibrium with JOREK and comparing it to the results
of another equilibrium solver allows to check some parts specific for $n=0$ like
poloidal field-coil contributions (see Sec.~\ref{:impl:boundint}). We are considering
an ITER-like limiter plasma. As seen from Figure~\ref{fig:equil-compare}, flux
surfaces and q-profile of the JOREK equilibrium agree very well with the
equilibrium computed by the CEDRES++ code~\cite{Hertout2011}.
Slightly different discretizations of the poloidal field coils might be responsible
for small remaining differences.

\subsection{Tearing Mode}

\begin{figure}
\centering
  \includegraphics[width=0.47\textwidth]{\figurefolder 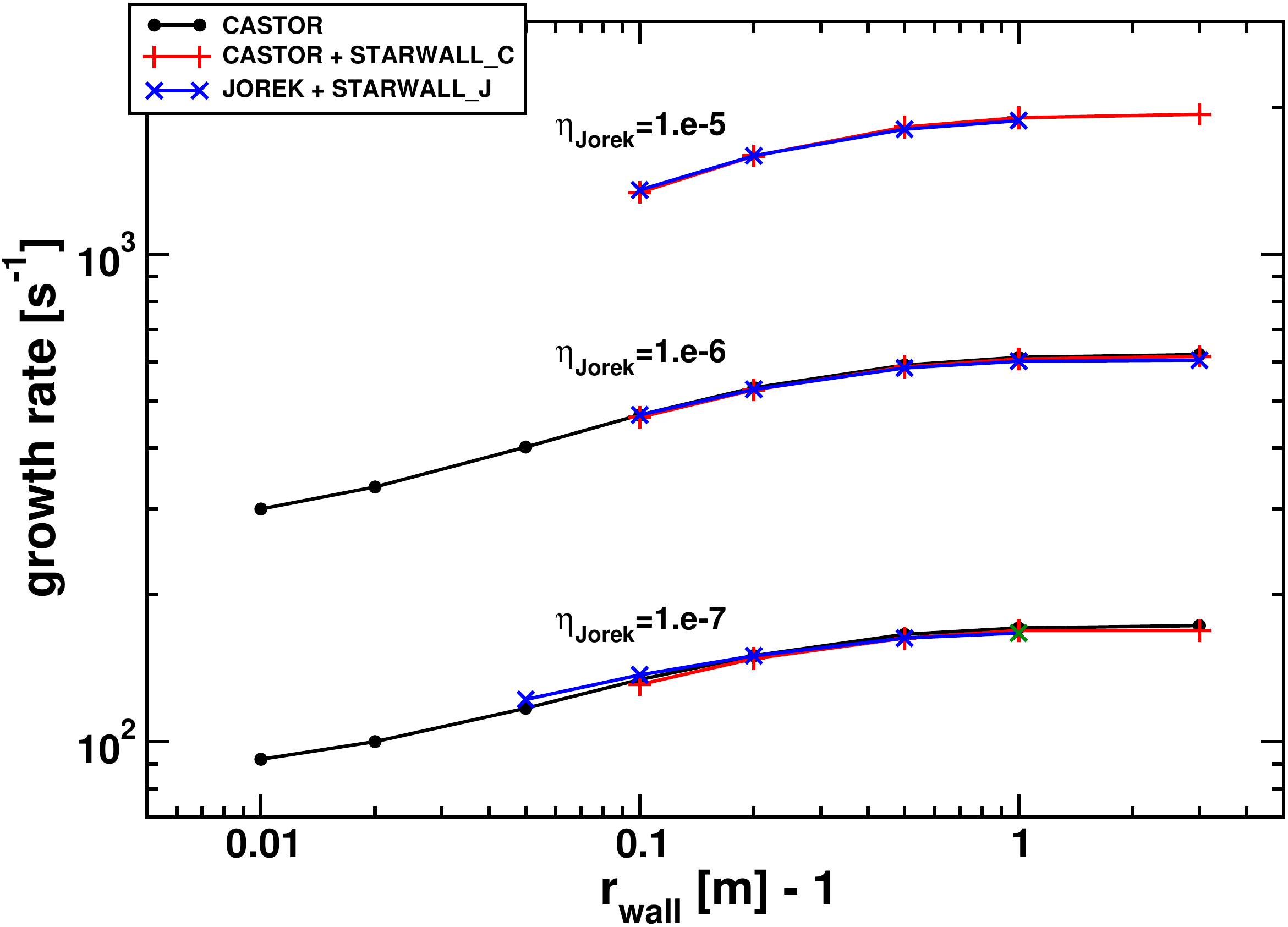} \hspace{5mm}
  \includegraphics[width=0.47\textwidth]{\figurefolder 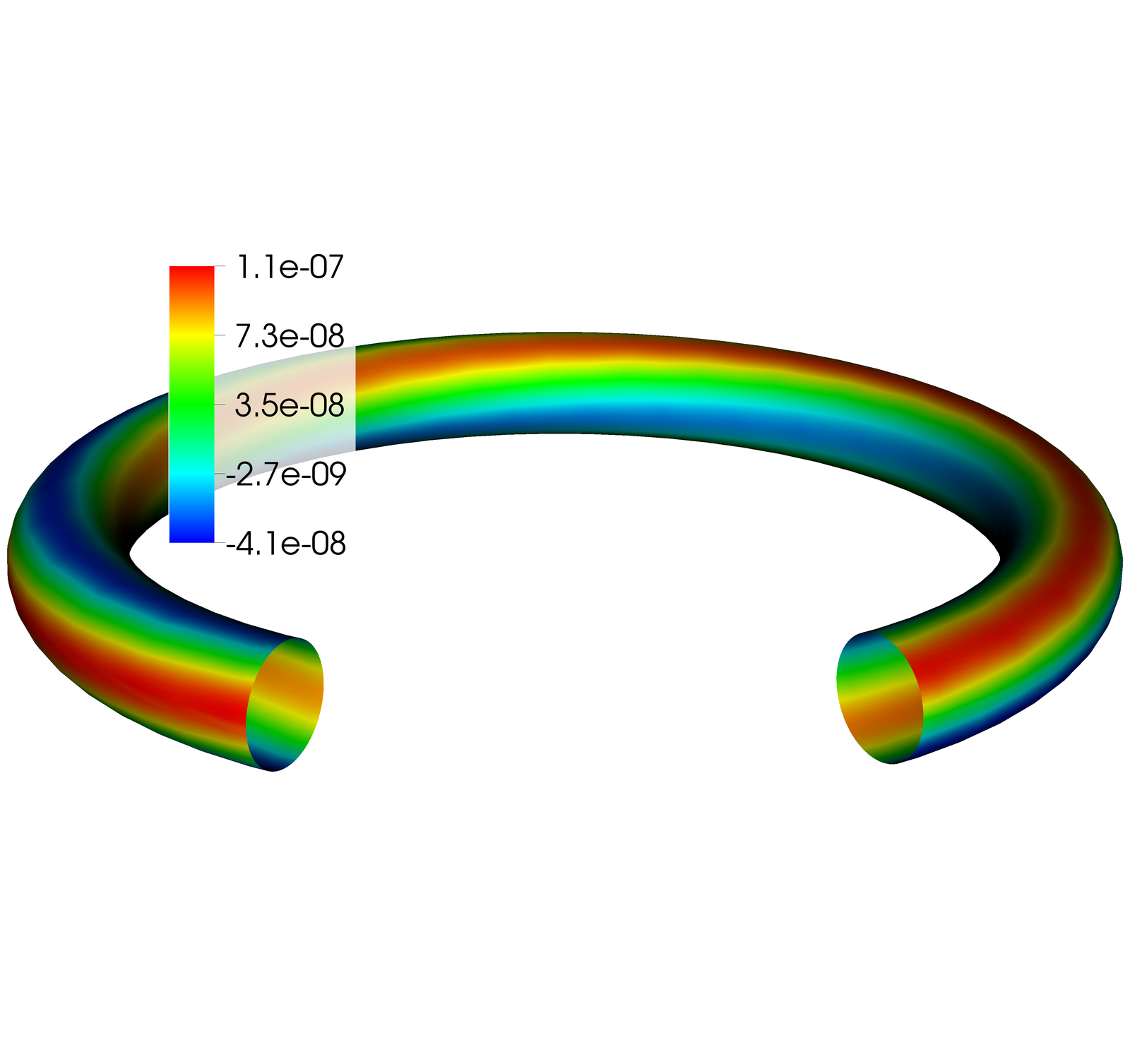}
\caption{
  \newline \textbf{Left:} Linear growth-rates of a tearing-mode in a circular plasma
  with concentric ideal wall are compared between \inred{CASTOR using its own vacuum field
  module, CASTOR coupled with STARWALL, and JOREK coupled with STARWALL}. Very
  good agreement is observed for a variety of different plasma
  resistivities and plasma-wall distances.
  \newline \textbf{Right:} The current potential distribution (arbitrary units)
  reflecting the $2/1$ tearing mode structure is plotted during the linear phase
  ($\eta_\text{Jorek}=1\cdot10^{-6}$, $r_\text{w}=1.2$).}
\label{fig:tearing-benchmark}
\end{figure}

A $2/1$ tearing mode in a circular plasma
with major radius $R=10$, minor radius $a=1$
and uniform plasma resistivity surrounded by an ideally conducting wall
is considered.
Figure~\ref{fig:tearing-benchmark} shows excellent agreement between JOREK
and the linear CASTOR code~\cite{Kerner1998}. Repeating the simulation for a
resistive wall with zero resistivity is physically identical, of course,
but numerically treated in a different way. The exact agreement we get here
proves consistency.

\subsection{Resistive Wall Mode}

\begin{figure}
\centering
  \includegraphics[width=0.48\textwidth]{\figurefolder 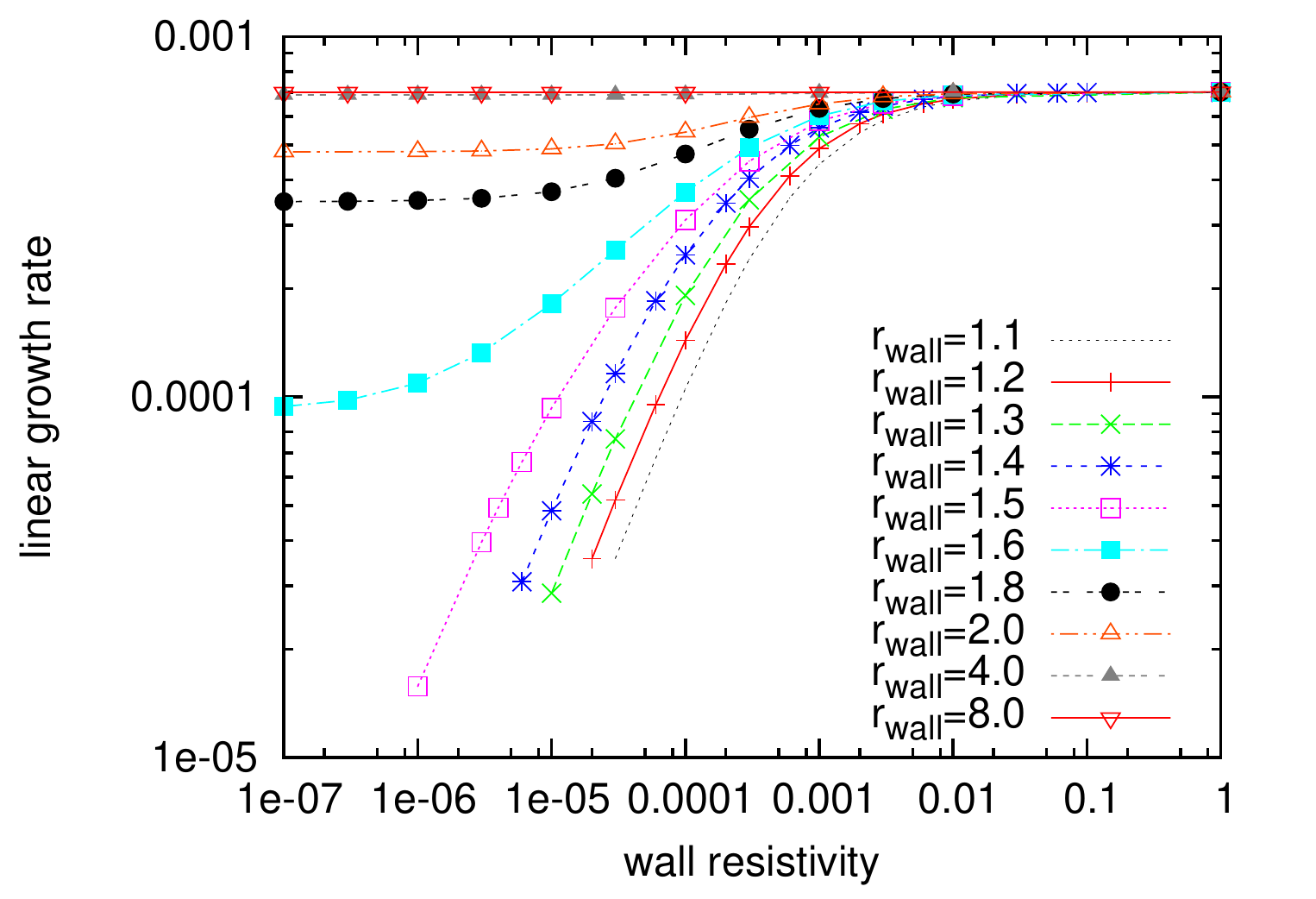} \hspace{2mm}
  \includegraphics[width=0.48\textwidth]{\figurefolder 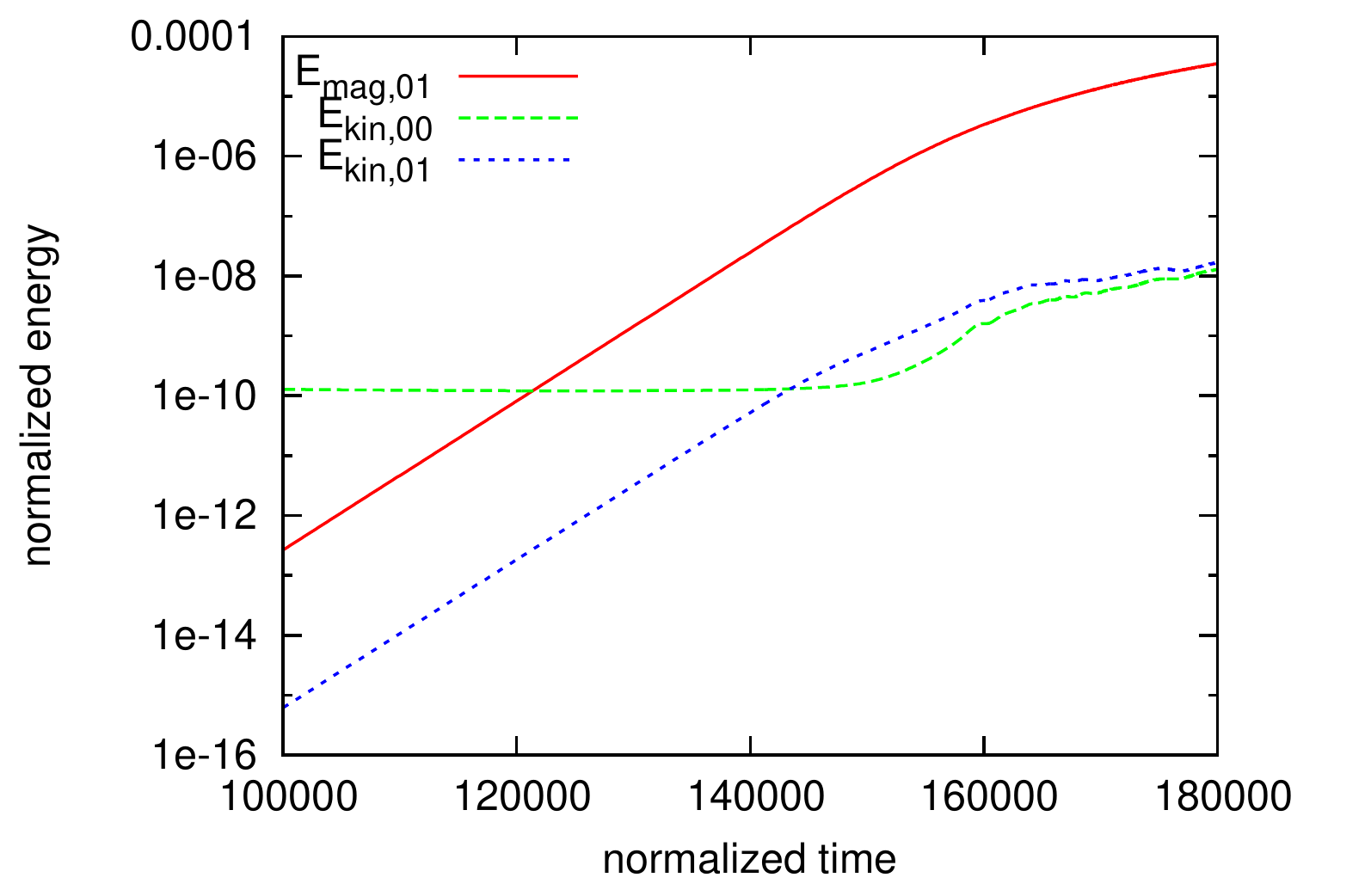}
\caption{\newline
  \textbf{Left:} Linear RWM growth rates extracted from the simulations are
  shown for a variety of wall radii and wall resistivities.\newline
  \textbf{Right:} Energy time-traces are shown for linear and saturation
  phase of an RWM ($r_\text{w}=1.2$, $\eta_\text{wall}=1\cdot10^{-4}$).
  In saturation, wall currents on the high-field side have an
  amplitude comparable to the low-field side while they are smaller in the linear
  phase.}
\label{fig:rwm}
\end{figure}

In this Section, results for $2/1$ resistive wall modes~(RWMs) in a circular
plasma with minor radius $a=1$ and major radius $R=10$ surrounded by a resistive
wall are presented. The left part of Figure~\ref{fig:rwm} shows linear
growth rates obtained for different wall radii $r_\text{w}$ as a function of the wall
resistivity. Clearly, the modes cannot be stabilized beyond a certain wall
distance from the plasma even at very low wall resistivities. The right part of the
Figure shows energy time-traces during the linear and saturation phase of a typical
simulation.

\section{Conclusions and Outlook}\label{:concl}

The ongoing implementation and verification of a resistive wall model in
the non-linear MHD-code JOREK was summarized. Benchmarks for a freeboundary
equilibrium and tearing mode cases show good agreement. Simulations of the
linear and non-linear phase of RWMs were presented (comparison to analytical
theory and linear codes ongoing).

Benchmarking will be continued and extended to non-linear comparisons. Realistic X-point
geometries will be considered requiring a special treatment of grid corners.
After completion, the code can be applied to a variety of MHD instabilities
interacting with conducting structures like resistive wall modes, edge localized modes,
vertical displacement events or disruptions.

\section*{Acknowledgements}

Simulations were mostly carried out on the HPC-FF computing cluster in J\"ulich,
Germany, and on the IFERC-CSC Helios computing cluster in Rokkasho-Mura, Japan.

\section*{References}

\bibliography{mybib}

\providecommand{\newblock}{}
\begin{thebibliography}{10}
\expandafter\ifx\csname url\endcsname\relax
  \def\url#1{{\tt #1}}\fi
\expandafter\ifx\csname urlprefix\endcsname\relax\def\urlprefix{URL }\fi
\providecommand{\eprint}[2][]{\url{#2}}

\bibitem{Chu2010}
Chu M~S and Okabayashi M 2010 {\em Plasma Physics and Controlled Fusion\/} {\bf
  52} 123001

\bibitem{Igochine2012}
Igochine V 2012 {\em Nuclear Fusion\/} {\bf 52} 074010

\bibitem{Czarny2008}
Czarny O and Huysmans G 2008 {\em J. Comput. Phys.\/} {\bf 227} 7423 -- 7445

\bibitem{Huysmans2010}
Huysmans G, Pamela S, Beurskens M, Becoulet M and van~der Plas E 2010 {\em
  Proceedings of the 23rd {IAEA} Fusion Energy Conference\/} (Daejon, South
  Korea) {THS/7-1}

\bibitem{Merkel2006}
Merkel P and Sempf M 2006 {\em Proceedings of the 21st {IAEA} Fusion Energy
  Conference\/} (Chengdu, China) {TH/P3-8}
  \urlprefix\url{www-naweb.iaea.org/napc/physics/FEC/FEC2006/papers/th\_p3-8.pdf}

\bibitem{Hoelzl2012A}
H\"olzl M, G\"unter S, Wenninger R, M\"uller W~C, Huysmans G, Lackner K, Krebs
  I and {the ASDEX Upgrade Team} 2012 {\em Physics of Plasmas (accepted)\/}

\bibitem{Strumberger2011A}
Strumberger E, Merkel P, Tichmann C,  and G\"unter S 2011 {\em Proceedings of
  the 38th {EPS Conference on Plasma Physics}\/} (Strasbourg, France) p5.082
  \urlprefix\url{http://ocs.ciemat.es/EPS2011PAP/pdf/P5.082.pdf}

\bibitem{HIRSCH1989}
Hirsch C 1989 {\em Numerical Computation of Internal and External Flows, Volume
  1, Fundamentals of Numerical Discretization\/} (Wiley) ISBN 978-0-471-92385-5

\bibitem{Hertout2011}
Hertout P, Boulbe C, Nardon E, Blum J, Br\'emond S, Bucalossi J, Faugeras B,
  Grandgirard V and Moreau P 2011" {\em Fusion Engineering and Design\/} {\bf
  86} 1045--1048 {Proceedings} of the 26th Symposium of Fusion Technology
  ({SOFT}-26)

\bibitem{Kerner1998}
Kerner W, Goedbloed J, Huysmans G, Poedts S and Schwarz E 1998 {\em J. Comput.
  Phys.\/} {\bf 142} 271 -- 303

\end{thebibliography}

\inred{%

\appendix

\section{Relation between ideal and resistive response matrices}\label{::app1}

Between the ideal and resistive response-matrices, the relation
\begin{equation}\label{eq:freebound220}
  \hat{M}^\text{id}_{i,j} \equiv \hat{M}^\text{ee}_{i,j} - \sum_k\hat{M}^\text{ey}_{i,k}\;\hat{M}^\text{ye}_{k,j}
\end{equation}
must hold which becomes obvious when letting $\eta_\text{w}\rightarrow0$ and
inserting Eq.~\eqref{eq:freebound200} into Eq.~\eqref{eq:freebound100}.
Also, the no-wall response can easily be identified as
\begin{equation}\label{eq:freebound240}
  \hat{M}^\text{nw}_{i,j} \equiv \hat{M}^\text{ee}_{i,j}
\end{equation}
when letting $Y_k\rightarrow0$. Eqs.~\eqref{eq:freebound220}
and~\eqref{eq:freebound240} can be used to cross-check
some parts of the implementation as both forms of the wall-treatment with
eliminated wall-currents (Eq.~\eqref{eq:freebound80}) and explicitly treated
wall-currents (Eqs.~\eqref{eq:freebound100} and~\eqref{eq:freebound200})
are implemented in JOREK.

\section{Magnetic field tangential to the interface}\label{::app2}

The magnetic field in the JOREK reduced-MHD model is given by
\begin{equation}\label{eq:app2:1}
  \mathbf{B}=\frac{F_0}{R}\hat{\mathbf{e}}_\phi+\frac{1}{R}\nabla\Psi\times\hat{\mathbf{e}}_\phi.
\end{equation}
Its component tangential to the boundary of the JOREK computational
domain (in the poloidal plane) can be determined from
\begin{equation}
  B_\text{tan}=(\mathbf{B}\times\hat{\mathbf{n}})\cdot\hat{\mathbf{e}}_\phi,
\end{equation}
where $\hat{\mathbf{n}}$ denotes the unit normal vector of the interface
and $\hat{\mathbf{e}}_\phi$ the normalized toroidal basis vector. Inserting
Eq.~\eqref{eq:app2:1} and using vector identities, this can be written in
the form
\begin{equation}\begin{split}
  B_\text{tan}
    &=-\frac{1}{R}\hat{\mathbf{e}}_\phi\cdot\left[\hat{\mathbf{n}}\times
      (\nabla\Psi\times\hat{\mathbf{e}}_\phi)\right] \\
    &=-\frac{1}{R}\hat{\mathbf{e}}_\phi\cdot\left[
      \underbrace{(\hat{\mathbf{n}}\cdot\hat{\mathbf{e}}_\phi)}_{\equiv0}\nabla\Psi
      -(\hat{\mathbf{n}}\cdot\nabla\Psi)\hat{\mathbf{e}}_\phi
      \right] \\
    &=\frac{1}{R}\hat{\mathbf{n}}\cdot\nabla\Psi,
\end{split}\end{equation}
which can be identified in the boundary integral of Eq.~\eqref{eq:freebound1100}
and replaced by the STARWALL response.
}

\end{document}